\documentclass[sigconf,nonacm]{acmart}

\usepackage{graphicx} 

\newcommand{\sys}{E-FuzzEdge}

\title{\sys: Optimizing Embedded Device Security with Scalable In-Place Fuzzing}
\author{Davide Rusconi}
\email{davide.rusconi@unimi.it}
\affiliation{%
  \institution{Università degli Studi di Milano}
  \city{Milan}
  \country{Italy}
}
\author{Osama Yousef}
\email{ossama.yousef@studenti.unimi.it}
\affiliation{%
    \institution{Università degli Studi di Milano}
    \city{Milan}
    \country{Italy}
}

\author{Mirco Picca}
\email{mirco.picca@unimi.it}
\affiliation{%
    \institution{Università degli Studi di Milano}
    \city{Milan}
    \country{Italy}
}

\author{Toffalini Flavio}
\email{flavio.toffalini@rub.de}
\affiliation{%
  \institution{Ruhr University Bochum}
  \city{Bochum}
  \country{Germany}
}

\author{Andrea Lanzi}
\email{andrea.lanzi@unimi.it}
\affiliation{%
  \institution{Università degli Studi di Milano}
  \city{Milan}
  \country{Italy}
}
\date{}

\begin{abstract}
In this paper we show \sys, a novel fuzzing architecture targeted towards improving the throughput of fuzzing campaigns in contexts where scalability is unavailable.  \sys addresses the inefficiencies of hardware-in-the-loop fuzzing for microcontrollers by optimizing execution speed. We evaluated our system against both real-world embedded libraries and state-of-the-art benchmarks, demonstrating significant performance improvements. A key advantage of \sys architecture is its compatibility with other embedded fuzzing techniques that perform on device testing instead of firmware emulation. This means that the broader embedded fuzzing community can integrate \sys into their workflows to enhance overall testing efficiency.
\end{abstract}

\begin{document}

\maketitle

\section{Introduction}

In the current digital era, computing extends beyond the traditional framework of desktops for daily activities and servers for demanding tasks. Instead, embedded devices are increasingly utilized in multiple industries, such as transportation, industrial automation, healthcare, and consumer electronics. Given their inherent resource limitations and the fact that they operate within safety-critical hard real-time environments, optimizing time and spatial efficiency is essential; however, ensuring security is of utmost importance. The increasing number of cyberattacks on embedded systems highlights the security challenges associated with the Internet of Things (IoT) and emphasizes the need for robust and systematic testing strategies. This deficiency does not stem from a lack of interest from the industry or academia but rather from the complex nature of embedded systems. Their architecture is vastly different from standard computing systems, making it difficult to apply general-purpose analysis tools effectively. The main issues include stringent resource limitations that hinder thorough analysis and the absence of typical features associated with general-purpose computers, such as filesystems, processes, or an operating system. 
Despite these limitations, the growing prevalence of embedded devices in safety-critical and resource-constrained environments has intensified the need for automated vulnerability testing techniques. However, testing methodologies developed for traditional computing platforms often fail to translate effectively to embedded environments due to their unique architectural constraints.

In response to these challenges, recent years have seen various efforts to adapt fuzzing techniques, one of the most effective automated vulnerability detection methods, to embedded systems. These adaptations fall primarily into two broad categories: (1)~\textit{Emulation-based techniques}, which attempt to execute firmware in a virtualized or rehosted environment, allowing a more scalable and introspective fuzzing process. (2)~\textit{In-place fuzzing}, which executes test cases directly on the embedded device, ensuring interaction with actual hardware peripherals and reducing false positives and negatives. Each approach presents unique trade-offs. Emulation-based techniques (e.g., rehosting and firmware transplantation) allow researchers to efficiently analyze firmware across multiple devices without requiring physical hardware. These methods enable fine-grained introspection, aiding in the detection of subtle software vulnerabilities. However, their effectiveness is highly dependent on the accuracy of peripheral modeling, a fundamental challenge, since embedded devices often include proprietary, undocumented, or highly customized hardware components. Inaccurate peripheral modeling can lead to incomplete or misleading security assessments.

In contrast, in-place fuzzing provides a more realistic security evaluation by running test cases on the actual embedded device, interacting with real peripherals, and uncovering hardware-related vulnerabilities that emulation might miss. This approach minimizes false positives and negatives and ensures a more accurate assessment of the behavior of the system under attack conditions. However, it introduces significant challenges, particularly in terms of scalability and efficiency. Due to severe resource constraints, the in-place fuzzing must operate within strict memory, processing, and power limitations. Additionally, communication overhead between the test system and the embedded device can dramatically impact performance, requiring careful optimization to ensure a viable fuzzing rate. Instances of these systems include IPEA~\cite{ipea} and $\mu$AFL, which exemplify hardware-in-the-loop methodologies operating directly on peripherals. IPEAFuzz employs minimalistic instrumentation to streamline data communication between the hardware and the fuzzer. In contrast, AFL exploits ARM-specific debugging capabilities for device interaction, eliminating the need for firmware instrumentation. 

Our research seeks to enhance the scalability and performance of these solutions by sustaining AFL++ technology while remaining orthogonal to the current state-of-the-art approaches. In particular, we introduce \textit{E-FuzzEdge}, an in-place fuzzer designed to enhance execution efficiency and provide a scalable resource-sensitive testing framework tailored to embedded environments.  Our project focuses on two main goals: first, to investigate methods for enhancing in-place fuzzing, thereby increasing the execution rate on the test device; second, to provide an open-source fuzzer to the community, which leverages cutting-edge technologies by adapting AFL++ for the IoT environment. 
In pursuit of these goals, we first developed a new fuzzing framework aimed at optimizing the use of embedded devices. This is achieved by minimizing the data transfer between the device that runs the inputs and the desktop responsible for handling feedback and producing new inputs. Our empirical observations identified this data exchange as the main source of overhead in hardware-in-the-loop fuzzing methods.
Furthermore, we designed a parallel system that enables uninterrupted execution while desktop instances produce new inputs, aiming to increase the execution frequency by cutting down on the downtime typically encountered when the fuzzer gathers and analyzes the feedback data after the execution of the test case. We then created a prototype of this system that uses AFL++, which is freely accessible, refining the current remote fuzzing technology available while maintaining its original architectural functions as much as possible. Our prototype was assessed in comparison to two leading-edge in-place embedded fuzzers: IPEA \cite{ipea} and $\mu$AFL \cite{mufuzz}. We noted a notable enhancement in execution speed, which illustrates that the proposed architecture has the potential to considerably boost the throughput of a fuzzing campaign. Moreover, the contributions of our model are orthogonal to both IPEA and $\mu$AFL meaning that a similar architecture can be integrated into these projects to further improve execution speed while retaining their key contributions to the topic of embedded fuzzing.
\section{Background}
\subsection{Fuzz Testing}

Fuzz testing, or \textit{fuzzing}, is an automated software testing technique introduced in the late 1980s~\cite{miller_fuzzing}. Over the years, it has become one of the most effective methods for discovering software vulnerabilities, contributing to the detection of numerous security flaws in widely used systems. The core principle behind fuzzing is to generate a large number of test inputs, either randomly or through systematic mutation, and to monitor the execution of the target program for unexpected behaviors, such as crashes, hangs, or incorrect outputs.

Architecturally speaking, modern fuzzers are composed of different parts as described by Fioraldi et al. \cite{LibAFL}: the input, the corpus that collects multiple inputs alongside their metadata, a scheduler that selects the next input to provide to the program, one or more stages that operate on the inputs that will be fed to the software, observers collecting data about a single execution, an executor tasked with running the specified program with the specified input, a feedback that classifies the executed inputs as interesting based on the observers' output, one or more mutators that, given one or more inputs, derive a new input, and a generator to create test cases from scratch.

Fuzzers can be classified according to various criteria. An important distinction lies in how they generate inputs. \textit{Generational fuzzers} create test cases from scratch using knowledge of input structure, such as grammars or protocol specifications, while \textit{mutational fuzzers} modify existing inputs through techniques like bit flips or byte alterations.

Another key distinction is the level of insight that a fuzzer has into the target program. \textit{Black-box fuzzers} operates without any internal knowledge of the system under test, relying purely on external observation of program behavior. \textit{White-box fuzzers}, on the other hand, has full access to the program’s source code or binary, leveraging advanced program analysis techniques to guide input generation. \textit{Grey-box fuzzers} represents an intermediate approach, using lightweight instrumentation to provide feedback on program execution, helping to improve coverage while maintaining efficiency.

Fuzzing techniques also differ in their execution models. One of the most widely used methods, employed in AFL++\cite{aflpp}, is the \textit{forkserver-based approach}, where the fuzzer spawns a new process for each test case to ensure isolation and reproducibility. Although effective, this method incurs overhead due to frequent process creation. An alternative is \textit{in-process fuzzing}, where the fuzzer and the target application share the same process space. This significantly improves execution speed but introduces robustness challenges, as unintended state changes may persist between test cases.

Architecturally speaking, modern fuzzers are composed of different parts, as described by Fioraldi et al. \cite{LibAFL}. A standard \emph{forkserver} greybox fuzzer such as AFL++\cite{aflpp} usually contains an executor that runs the program under test by forking it, an observer that obtains coverage information for the tested input, a feedback system that, based on the coverage data, decides whether the input is interesting, multiple mutators that modify the inputs to expand the corpus, i.e. the set of inputs to feed the program, and a scheduler that chooses which input to feed the process. 
\subsection{Fuzzing in Embedded Systems}

Although fuzzing has been widely successful in conventional software security, applying it to embedded systems presents unique challenges. Unlike general-purpose computing environments, embedded devices often have stringent constraints on memory, processing power, and energy consumption. Many lack fundamental system components such as filesystems, process isolation, or even complete operating systems. These limitations make direct adoption of traditional fuzzing techniques impractical.

To address these constraints, researchers have developed two main approaches for fuzzing embedded devices. The first is \textit{emulation-based fuzzing}, which executes firmware within a virtualized environment, allowing for large-scale testing and deep introspection~\cite{fuzzware, p2im, halucinator}. This method benefits from scalability, as it enables security analysts to test the firmware independently of the physical device. However, its effectiveness is highly dependent on the accuracy of peripheral modeling. Since embedded systems often incorporate proprietary or undocumented hardware components, emulation inaccuracies can lead to misleading results or missing vulnerabilities.

In contrast, \textit{in-place fuzzing} executes test cases directly on the embedded device, ensuring interaction with real peripherals and capturing vulnerabilities that might not manifest in an emulated environment~\cite{mufuzz, ipea}. This approach reduces the risk of false positives and negatives, providing a more realistic security assessment. However, it introduces significant challenges related to scalability and efficiency. Since embedded systems typically lack sufficient computational resources, fuzzing must be carefully optimized to minimize memory usage, CPU overhead, and power consumption. Additionally, communication between the fuzzer and the device can create bottlenecks, further limiting execution speed.

Given these trade-offs, achieving an effective fuzzing solution for embedded systems requires balancing the scalability and introspection capabilities of \textit{emulation-based approaches} with the accuracy and realism of \textit{in-place methods}. This work introduces \textit{E-FuzzEdge}, an optimized fuzzer in place that improves execution efficiency through parallelization and streamlined communication, enabling scalable and effective vulnerability detection in embedded environments.

\section{\sys~Design}
The main focus of our project is designing an efficient greybox in-place fuzzing architecture for embedded devices, minimizing execution overhead while maintaining adaptability across different hardware environments.
The core requirements of our architecture are versatility and efficiency. We can examine the former along two dimensions: the adaptability of the communication channel to suit the diverse environments typical of embedded devices, and a decrease in platform requirements, such as processes and filesystem. This adjustment is necessary for platforms with extremely simplistic or non-existent operating systems.
In terms of efficiency, the main issue is the additional burden caused by the communication channel. As this burden is intrinsic to the platform, our method emphasizes reducing its impact instead of trying to remove it altogether. We address this overhead by employing two main strategies: decreasing the volume of data transmitted through the communication channel and reducing device idle time through enhanced data transmission and processing.

\begin{figure}
    \centering
    \includegraphics[width=0.95\linewidth]{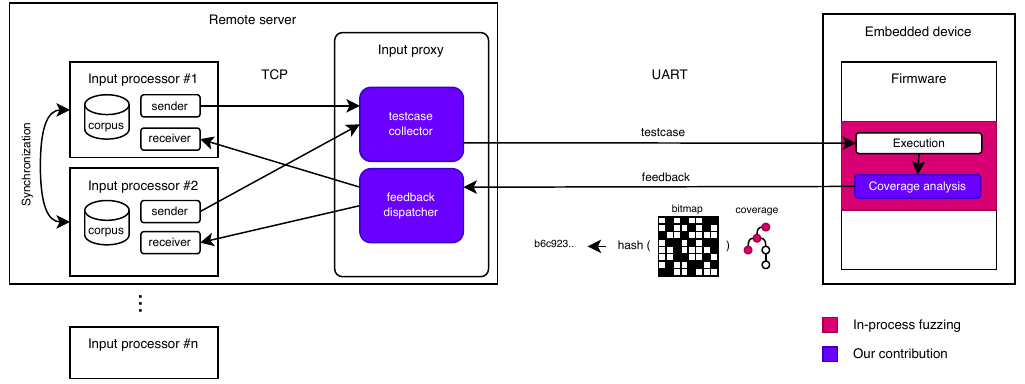}
    \caption{\sys\ Architecture}
    \label{fig:arch}
\end{figure}

\subsection{Architectural Model}
We devised our architecture in order to satisfy the mentioned requirements, particularly the one related to overhead, since versatility is related to the implementation and not to the model.
Our system remains consistent with standard forkserver-based fuzzing designs, but we decouple input processing,mutation, and scheduling, from test case execution and feedback collection. 
Figure \ref{fig:arch} shows the three elements upon which our system is built: multiple \emph{input processors}, a \emph{input proxy}, and a single \emph{input executor}

\paragraph{Input Processor}
Multiple processors run on a desktop device, leveraging superior computing resources and filesystem access to enhance usability and accelerate operations that do not require direct interaction with the embedded hardware. 
They are tasked with all operations related to the processing of test cases, such as scheduling and mutation, based on coverage feedback. The choice of using multiple \emph{input processors} stems from the idea of maximizing the use of the device. As we said, sending and retrieving data on the communication channel is a slow process; the use of a one-to-many architecture ensures that the single executor does not remain idle while a processor is retrieving and analyzing the data, but runs the inputs provided by the other executors, thus maximizing the throughput.

\paragraph{Input proxy}
A lightweight middleware sitting between processors and devices.
It exposes a stable host-facing interface and adapts to the device link (TCP or UART).
For UART it handles practical details like DMA cyclic reception and padding so half/full-buffer interrupts fire predictably. 
The choice of a proxy comes with the benefit of easing the process of changing the communication protocol between the desktop system and the embedded device, which is particularly relevant in a highly heterogeneous context in terms of communication protocols.

\paragraph{Input Executor}
The executor runs in-process with the target firmware and a small harness without relying on processes or fork() that can be not available when fuzzing bare-metal.
The harness is tasked with initializing the necessary peripherals for testing the firmware, retrieving inputs from the proxy and feeding them to the program when needed. 
The execution of test cases follows the standard fuzzing loop of execution and feedback analysis but is a persistent process, meaning that the system is not rebooted after each testcase. While this approach can create issues due to separate tests concurring in the stimulation of a bug. We argue that this issue can be mitigated by careful harness design and the benefit in terms of executions per second makes adopting the persistent approach worth it.
The executor is also tasked with processing coverage data related to the executed test; this allows us to minimize the data transmitted on the communication channel, in the form of a flag indicating whether new bits were found and a checksum, and not perform a full-bitmap transmission that would significantly hinder fuzzing performance.

\subsection{Communication Protocol}

The protocol mirrors the classic fuzzing loop of schedule, mutate, execute, evaluate, but is engineered to minimize bytes in flight and device idle time. During initialization, processors establish TCP sessions with the proxy, the proxy opens or attaches to the device link, and the executor configures its peripherals and allocates the coverage structures it will maintain during the campaign. Any static configuration that affects message formats, such as the effective map size or maximum payload, is broadcast from the proxy to processors so formatting remains consistent.

In steady state each processor prepares a test case and submits it to the proxy; the proxy forwards the frame to the device ; the executor runs the harnessed target to completion or timeout, computes coverage locally, and performs on-device triage against the cumulative map. It then sends a compact feedback record composed of a fault code when applicable, a boolean flag indicating whether new bits were observed, and, if enabled, a short checksum. The proxy routes this record back to the originating processor, which decides whether to retain or discard the input and immediately schedules the next one. Because only minimal feedback travels on the hot path, steady-state traffic remains small and predictable.

Transport specifics are handled transparently by the proxy.  Over UART, the executor configures DMA in cyclic mode and relies on half and full-buffer interrupts to delimit frames, while the proxy pads headers and bodies so those interrupts fire deterministically. 

\section{Implementation}
In order to test our model, we developed a prototype based on AFL++ \cite{aflpp}. Although AFL++ already provides a proxy system, we developed the remote fuzzing system from scratch because that system allows only a one-to-one fuzzing architecture and requires the fuzzer to send back the whole coverage map, significantly slowing down the whole process. Additionally, we modified the forkserver to allow remote communication with the embedded device and adapted the feedback collection mechanism to process the input without the need for the coverage map, since the task of analyzing the input is performed by the input executor. Moreover, we modified the compiler runtime to accomplish two objectives: first, we reduced the data needed for the fuzzing campaign in order to correspond to the strict space requirements present within the embedded context; second, we added the infrastructure necessary to initialize the peripherals needed for communicating with the proxy and retrieve input data from it. Lastly, we developed the proxy system to mediate between the afl instances and the embedded system. In total, we contributed 6534 lines of C/C++ code. 

\section{Evaluation}

We evaluate \sys\ in two settings: a desktop-only setup where input processors and the executor communicate over TCP and a hardware-in-the-loop setup using an STM32L053R8 board as the embedded device and a desktop host (Intel i7-9750H, 16~GB RAM) performing seed processing. In the embedded setup, host–device communication uses UART.

For each firmware in our dataset, we ran an average of seven fuzzing campaigns of 4~hours each for $n\in\{1,2,3,4\}$ parallel input processors. We report executions per second (exec/s) aggregated over each campaign. The 4-hour duration allows the fuzzer to stabilize while keeping runs practical.

\subsection{Dataset}
On desktop we used \texttt{xpdf}, a standard fuzzing target. For the assessment in the embedded context evaluation we used a dataset comprised of 10 open source embedded firmwares. Among these two were taken from the OAT dataset \cite{oat}, whereas the remaining 8 were taken from well-known open source
repositories—Arduino Projects~\cite{arduino} and Raspberry Pi Pico Examples~\cite{picopi}.
Each embedded firmware was slightly modified to remove the original \texttt{main} and to install a simple, in-process harness (standard practice in fuzzing). Common input functions (e.g., \texttt{fgetc}, \texttt{scanf}) were reimplemented to read from the proxy-provided test case rather than \texttt{stdin} or UART, leaving application logic unchanged.

\subsection{Results}
\paragraph{Desktop Setting}

To bound the potential benefits of our host–side concurrency model under a fast link and a high-performance executor, we replicated the client–proxy protocol over TCP on a desktop host and executed the target with a single forkserver-backed \emph{input executor}. With one executor, baseline throughput is \textbf{20.24 exec/s}. Allowing multiple executors in parallel yields speedups of \(2.09\times\), \(2.10\times\), \(3.20\times\), \(5.11\times\), and \(6.81\times\) for \(\{2,3,4,6,8\}\) executors, corresponding to absolute throughputs of \(\{42.28, 42.44, 64.69, 103.45, 137.96\}\) exec/s. It is clear that as concurrency increases, scaling trends toward CPU-bound behavior: improvements persist until available cores and memory bandwidth saturate, at which point speedup becomes sub-linear.

This configuration is a stress test rather than a recommended desktop setup: general-purpose binaries can already fork freely without our proxy, and running multiple independent fuzzers per core is standard practice; on typical desktop targets, execution rates are \emph{on the order of thousands of exec/s}, i.e., at least an order of magnitude higher than our proxy-mediated desktop measurements even after adding parallelism. Nevertheless, the experiment is informative for embedded scenarios because it isolates the effect of \emph{low-latency transport plus host-side concurrency} from device constraints. In other words, it approximates the leeway available to high-performance boards connected over fast links where the communication channel does not dominate. Two practical takeaways follow. First, when the link is fast enough, parallel producers convert directly into higher executions per second until the host CPU, not the channel, becomes the bottleneck. Second, the exact “knee” depends on the balance between per-input compute in the target and execution costs. This behavior mirrors our embedded results, where the knee appears earlier because the on-device executor and slower link saturate sooner. Overall, the desktop scaling curve provides an optimistic upper bound on the gains a high-performance board and transport could realize under the same one-executor/many-producer architecture.

\paragraph{Embedded Setting}

Table~\ref{tab:res} reports executions per second (exec/s) by firmware and by the number of parallel processors feeding the device. Values in parentheses are per-target speedups relative to one processor, and the “Total” row is the geometric mean speedup across firmwares for that column (targets without data are excluded).

The headline result is that adding a \emph{second} processor consistently improves throughput: the geometric-mean speedup across firmwares is \textbf{1.27$\times$}. A third processor yields \(\,1.06\times\) on average, and a fourth \(\,0.95\times\), indicating diminishing—and sometimes negative—returns beyond two. Concretely, 7/10 firmwares improve at two processors (typical gains \(\approx\)1.3–2.0\(\times\)), 2/10 are unchanged, and 1/10 regresses due to internal delays. This pattern reflects the intended division of labor: a second producer keeps the executor busy while the first is scheduling and processing feedback, but additional producers mostly contribute queueing and per-message overhead once the device and link are saturated.

Saturation is target-specific. The best-performing \(n\) is \(2\) for \texttt{discokeyboard}, \texttt{huemotion}, \texttt{rflock}, and \texttt{xml} (4/10); \(3\) for \texttt{ledmatrixpainter}, \texttt{thermostat}, and (effectively) \texttt{musiccontroller} (3/10); \(4\) for \texttt{modbus} (1/10); and \(1\) for \texttt{miniigstats} and \texttt{pixelpainter} (2/10), where fixed delays cap exec/s regardless of host-side concurrency. These outcomes align with intuition about where time is spent: when per-input device time and channel overhead dominate, extra processors add little beyond contention; when the target spends more time per input (e.g., protocol parsing in \texttt{modbus} or deeper control-flow in \texttt{thermostat}), a second processor converts directly into near-linear gains, with limited benefit thereafter. Operationally, on UART-class links and STM32L0-class boards, two processors are a strong default; the precise knee shifts with firmware behavior and transport latency/bandwidth, but the guiding principle remains to minimize bytes on the wire, keep the executor in a persistent loop, and scale concurrency on the host until the device or link becomes the bottleneck.

\begin{table*}[t]
\centering
\caption{Number of executions per second for each firmware}
\label{tab:res}
\scalebox{0.79}{
\begin{tabular}{l|c|c|c|c}
\toprule
Firmware name & 1 Instance & 2 Instances & 3 Instances & 4 Instances  \\
\midrule
discokeyboard   & 14.36 (1.00x) & 21.17 (1.47x) & 10.03 (0.70x) & 10.42 (0.73x) \\
huemotion       & 14.31 (1.00x) & 21.68 (1.52x) & 10.03 (0.70x) & 10.08 (0.70x) \\
ledmatrixpainter& 18.59 (1.00x) & 23.71 (1.28x) & 23.93 (1.29x) & 12.01 (0.65x) \\
miniigstats     & 7.41 (1.00x)  & 4.64 (0.63x)  & 4.67 (0.63x)  & 4.60 (0.62x)  \\
modbus          & 16.79 (1.00x) & 25.76 (1.53x) & 26.07 (1.55x) & 27.89 (1.66x) \\
musiccontroller & 0.94 (1.00x)  & 0.94 (1.00x)  & 0.95 (1.01x)  & 0.94 (1.00x)  \\
pixelpainter    & 2.62 (1.00x)  & 2.62 (1.00x)  & 2.61 (1.00x)  & 2.61 (1.00x)  \\
rflock          & 15.89 (1.00x) & 23.94 (1.51x) & 21.94 (1.38x) & 20.00 (1.26x) \\
thermostat      & 16.62 (1.00x) & 32.96 (1.98x) & 33.32 (2.00x) & 24.70 (1.49x) \\
xml             & 3.22 (1.00x)  & 4.17 (1.30x)  & -  & -  \\
\bottomrule
Total           & 1.00x         & 1.27x         & 1.06x         & 0.95x        \\
\hline
\bottomrule
\end{tabular}
}
\end{table*}



\section{Discussion}
Our evaluation has demonstrated the effectiveness of our parallel architecture, but it also highlighted several key limitations and avenues for future work.

\paragraph{Evaluation scope.}
As shown in the previous section, \sys\  improves fuzzing throughput when multiple input processors handle testcases in parallel. Our evaluation, however, is limited to a single MCU family (STM32L0) and two channels (TCP on desktop, UART on device). While we do not expect an inversion of the observed trend, more powerful boards and faster links will likely shift the saturation point beyond two processors. Replicating on additional MCUs and transports (e.g., USB CDC/SPI) is future work.

\paragraph{Memory overhead.}
Our prototype maintains one temporary coverage map per execution and one cumulative map per input-processor instance on the device. On memory-constrained boards with large targets, these allocations may be prohibitive. A straightforward extension is to use a single cumulative map shared across processors, reducing redundancy; this needs careful evaluation to rule out contention and other side effects.

\paragraph{Number of input processors.}
The current parallelism level is fixed at campaign start. Results indicate the optimal number of processors depends on both firmware and target hardware. On STM32L053R8, two processors are typically best. We plan to improve in this area by enabling runtime negotiation of the number of input executors; this approach can increase/reduce the number of input processors dynamically to keep the board saturated while minimizing the overhead caused by the communication channel in order to optimize the system throughput.

\paragraph{Crash Handling} Currently, we rely on timeouts to detect crashes. While this is a common and practical approach, it cannot distinguish a true crash from a simple timeout due to a delay in the target firmware's execution. This limitation can be addressed by implementing custom error handlers within the firmware to signal the host when a crash occurs. Furthermore, our architecture's high throughput, while beneficial, also exacerbates a core problem of in-place fuzzing: the need for frequent device resets upon crash detection. An increased crash discovery rate, while a sign of a successful fuzzing campaign, can lead to more frequent reboots, which in turn can significantly impact overall performance and limit the practical benefits of our architecture.

\section{Related Work}

Fuzzing techniques for embedded devices can be broadly classified into two major categories: hardware-in-the-loop approaches and emulation-based approaches.

{\bf Emulation-Based Approaches.} Emulation-based fuzzing techniques aim to execute firmware in a simulated environment, often without requiring access to the physical hardware. This approach enables scalable testing while avoiding hardware constraints, making it particularly useful for analyzing a wide range of embedded systems. Several notable works fall into this category. DICE \cite{dice}, Fuzzware \cite{fuzzware}, P2IM \cite{p2im}, and Laelaps \cite{laelaps} focus on peripheral modeling and inference. These frameworks leverage techniques such as symbolic execution, peripheral abstraction, and model learning to approximate the behavior of real hardware peripherals. This enables effective firmware execution in an emulated environment, improving the efficiency of fuzz testing. Other approaches focus on rehosting firmware in a way that abstracts the underlying hardware while preserving its execution semantics. HALucinator \cite{halucinator} targets the hardware extraction layer (HAL) by intercepting HAL calls and providing simulated or stubbed responses to maintain the firmware execution flow. Pretender \cite{pretender}, on the other hand, models Memory-Mapped I/O (MMIO) peripherals by dynamically observing and learning from the behavior of the original hardware. These techniques allow fuzzing frameworks to execute firmware more accurately in an emulated environment, while reducing the reliance on manually crafted peripheral models. Despite their advantages, emulation-based approaches often face challenges in accurately reproducing real-world execution environments, particularly when dealing with complex or undocumented peripherals. The inability to perfectly replicate hardware interactions can sometimes lead to discrepancies between the observed and actual firmware behavior.

{\bf Hardware-in-the-Loop Approaches.} In contrast to emulation, hardware-in-the-loop (HIL) fuzzing techniques execute firmware directly on the target device, allowing more accurate testing by interacting with actual hardware components. IPEAFuzz \cite{ipea} introduces a lightweight instrumentation method that facilitates data exchange between the target device and the fuzzer, allowing efficient real-time monitoring and feedback collection. $\mu$AFL \cite{mufuzz} employs ARM-specific debugging mechanisms, such as hardware failure points and trace features, to enable fuzzing without requiring modifications to the firmware itself. This approach minimizes interference while preserving the fidelity of the testing process. HIL approaches provide high-fidelity and realistic testing conditions, but suffer from hardware dependency and scalability issues. The requirement for physical devices imposes constraints on execution speed, automation, and reproducibility, making large-scale or parallel testing more challenging compared to emulation-based methods. 

Both hardware-in-the-loop and emulation-based approaches play crucial roles in fuzzing embedded systems, each with its own trade-offs. While emulation enables scalability and flexibility, it struggles with hardware fidelity. In contrast, hardware-in-the-loop methods offer high realism but face challenges related to scalability and accessibility. Recent advances, such as hybrid approaches that combine elements of both techniques, are emerging to bridge these gaps, offering promising directions for future research in embedded firmware security testing. In our work, we present a novel fuzzing framework that is leveraged to improve scalability and speed while maintaining high fidelity in firmware testing. Our approach outperforms existing methods by making it well-suited for large-scale embedded system security testing.

\section{Conclusion}

In this work we presented \sys, an embedded-system focused in-place fuzzer. With this work we studied how to improve the execution per second of a fuzzer and discovered that parallelizing the fuzzing instances even with a single executor on the embedded device can significantly improve the throughput of the system. We argue that the relevance of this result can be valid in and of itself since it can benefit already existing in-place embedded fuzzers such as IPEA Fuzz \cite{ipea} and $\mu$AFL \cite{mufuzz} given that this improvement is orthogonal to their contribution. Moreover we developed and distributed a prototype based on AFL++ thus bridging the gap between the state of the art in greybox desktop fuzzing and the tooling currently available in the embedded context.
\bibliographystyle{plain}
\bibliography{bibliography}

\end{document}